# Continuous Emotion Recognition during Music Listening Using EEG Signals: A Fuzzy Parallel Cascades Model

Fatemeh Hasanzadeh, Mohsen Annabestani, Sahar Moghimi

**Abstract**— A controversial issue in artificial intelligence is human emotion recognition. This paper presents a fuzzy parallel cascades (FPC) model for predicting the continuous subjective appraisal of the emotional content of music by time-varying spectral content of EEG signals. The EEG, along with an emotional appraisal of 15 subjects, was recorded during listening to seven musical excerpts. The emotional appraisement was recorded along the valence and arousal emotional axes as a continuous signal. The FPC model was composed of parallel cascades with each cascade containing a fuzzy logic-based system. The FPC model performance was evaluated by comparing with linear regression (LR), support vector regression (SVR) and Long-Short-Term-Memory recurrent neural network (LSTM-RNN) models. The RMSE of the FPC was lower than other models for estimation of both valence and arousal of all musical excerpts. The lowest RMSE was 0.089 which was obtained in estimation of valence of MS4 by FPC model. The analysis of MI of frontal EEG with the valence confirms the role of frontal channels in theta frequency band in emotion recognition. Considering the dynamic variations of musical features during songs, employing modeling approach to predict dynamic variations of the emotional appraisal can be a plausible substitute for classification of musical excerpts into predefined labels.

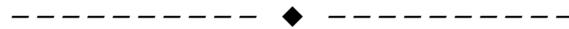

## 1 INTRODUCTION

Emotions play a fundamental role in human life. If the machine can recognize human emotions, they can work more efficiently because they will be more intelligent in dealing with human beings. Considering the importance of emotion recognition, this issue has drawn the attention of many researchers. To create machines that can recognize emotions, a great body of research has been dedicated to investigating the neural correlates of emotions. In these efforts, emotions have been elicited by different pictorial [1], musical [2-4] and video [5-7] stimuli. Music listening comprises a variety of psychological processes, e.g. perception and multimodal integration, attention, syntactic processing and processing of meaning information, emotion, and social cognition [8]. Music can create powerful emotions[9]. Emotional processing involves different structures of human brain and induces changes in their activity [8]. It also causes some other physiological responses which are secondary effect of brain activity, e.g. change in heart rate [10], skin conductance [11], and body temperature [11]. Different modalities including PET [12], fMRI [13], NIRS [14] and EEG [15] have been used by researchers to study neural correlates of emotions. High temporal resolution, portability, and relatively low cost of data recording has made EEG a suitable candidate for investigating the neural correlates of various cognitive functions, including emotion. During the past two decades EEG power spectra in different frequency bands namely theta (4–8 Hz), alpha (8–13 Hz), beta (13–30 Hz) and gamma (30–49 Hz) have been investigated in many studies to identify changes due to emotional processing [10]. Balconi et al. used EEG frequency bands along with hemodynamic measurements, to investigate affective responses in the brain [16]. Sammler et al. demonstrated that listening to pleasant music results in an increase in the theta power of EEG signals in the frontal midline region [10]. Higher frontal midline theta band energy for liked music was also observed by Balasubramanian et al. They have analyzed EEG component energy obtained by wavelet packet decomposition while listening to liked and disliked music [4]. Zheng used EEG spectral power as a feature for EEG channel selection and emotion recognition through sparse canonical correlation analysis [17]. Ozel et al. applied multivariate synchrosqueezing transform to extract EEG features for classifying emotional states. They reported 93% classification accuracy for one of the emotional states [18]. Hasanzadeh et al. used nonlinear autoregressive exogenous model and genetic algorithm to predict emotional state during listening to music by EEG power spectrum[19].

Some studies applied EEG band power alongside with facial expressions to compare their ability in emotion recognition [7, 20]. However Koelstra et al. reported more accurate emotion recognition by combination of EEG and face modalities [20] but Soleymani et al. founded superior

- *Fatemeh Hasanzadeh is with the Department of Electrical Engineering, K.N. Toosi University of Technology, Tehran, Iran. E-mail: f.hasanzadeh22@yahoo.com.*
- *Mohsen Annabestani is with the Department of Electrical Engineering, Sharif University of Technology, Tehran, Iran. E-mail: annabestany@gmail.com*
- *Sahar Moghimi is with the Department of Electrical Engineering, Ferdowsi University of Mashhad and with Rayan Center for Neuroscience and Behavior, Ferdowsi University of Mashhad, Mashhad, Iran. E-mail: s.moghimi@um.ac.ir.*

results by only facial expressions [7].

In addition to the spectral power of EEG signals, connectivity indices extracted from EEG networks during an emotional assessment of musical stimuli have been studied. Hasanzadeh et al. investigated connectivity of brain networks during listening to joyful, neutral and melancholic music using partial directed coherence. The results demonstrated that listening to joyful music increases intra-region effective connectivity in the frontal lobe [21]. Directed transfer function was another connectivity measure that was employed to study the effect of listening to emotional music on EEG networks [22]. Li et al. have analyzed EEG networks constructed by phase-locking value while watching music videos. They have investigated power spectral features, features based on network topology and their combination for emotion classification [23].

In most of the studies on emotion recognition, emotions have been described by a discrete emotion model including some specific adjectives or descriptors for basic emotions[1-3]. In these studies, emotion recognition was reduced to a classification problem. K-nearest neighbor [1, 2, 18], Support vector machine(SVM) [2, 3, 18, 23], classifier based on quadratic discriminant analysis(QDA) [1], and neural networks (NN) [2, 3, 24] are some of the classification techniques applied to date for the specific purpose of emotion classification. Choosing a limited set of emotion labels is not plausible for describing the emotional content of music since the intended adjective may not exist in the emotion descriptor set defined by the examiner [9]. The dimensional emotion model was employed to solve this problem. This model consists of emotional dimensions that encompass various emotions [9]. Russel proposed a two dimensions model for emotion-based on pleasure and activation which is referred to as valence and arousal [25]. A different dimensional model with two separate arousal dimensions namely energetic and tension arousal has been proposed by Thayer [26]. The dimensional model is not necessarily limited to two dimensions [9]. Some of the studies which applied dimensional model discretized the continuous space of emotion [4, 20], however it is reported that music can induce more than one emotion and using a discrete label may not reflect the emotional richness of music [9]. Therefore, a time-varying measure can be beneficiary for a more realistic quantification of emotional appraisal. Only a few studies applied continuous affect recognition which is done by using regression or neural network models. The models that are mostly used are linear regression (LR) [5, 7, 27], support vector regression (SVR) [7, 28], Long-Short-Term-Memory recurrent neural network (LSTM-RNN) [7, 28, 29]. Regression models are parametric models which have some pitfalls such as requiring prior knowledge of the system and low robustness to noise [30]. Neural networks have some defects, including massive computation, lack of interpretability for amending a specific behavior and difficulty in learning parameter setting [31]. In the other hand fuzzy systems are capable of describing system behaviors with simple conditional IF-THEN rules and are more interpretable systems [31]. Considering the advantages of fuzzy systems, it is applied in various applications such as image processing[32], blind source separation[33], descriptive educational evaluation[34], modeling of smart material behviour[35].

In the current study, we have proposed a model based on fuzzy systems for continuous emotions recognition. We are aimed to predict the time-varying emotions experienced by individuals during listening to music by EEG features. To investigate the dynamic nature of emotions a time-varying two-dimensional emotional rating paradigm was employed. Wavelet analysis was used to extract the time-varying power spectrum of EEG signals in theta, alpha, and beta frequency bands. To choose EEG features that are highly related to emotion, mutual information (MI) was applied. Our proposed model which is fuzzy-based exploits the extracted EEG features to predict the time course of the emotions expressed by individuals. This model enables us to go beyond the predefined labels and provides predictions of the time-varying emotional appraisal based on the neural correlates of the stimuli. In addition to the proposed model, LR, SVR and LSTM-RNN models are employed for continuous emotion recognition.

The main contributions of this work are as follows. First, we have proposed a novel model for emotion recognition for the first time. The superior performance of the proposed model is confirmed by comparing the model error with other models, including LR, SVR, and LSTM-RNN. Second, we have applied emotion in our analysis as a continuous signal, not fixed labels. Third as far as we know this the first study that investigated the MI of EEG features with emotion dimensions.

## 2 MATERIALS AND METHODS
### 2.1 Subjects

We recruited fifteen volunteers for this study. Participants were non-musicians (age: 21±3 yrs, 3 males) with a similar educational background (undergraduate or MSc students). All subjects were right-handed, had normal hearing, and no history of any neurological and psychological disorders. They also reported that they had normal nocturnal sleep (7 to 9 hours starting 10-12pm) one week before the experiment. The participants reported no use of caffeine, nicotine, and energy drinks 24 hours before the experiment, and had not excessive physical activity 24 hours before the recording session. The study protocol was approved by the Ethics Committee at the Ferdowsi University of Mashhad.

### 2.2 Stimuli
Stimuli were selected according to a previous study [36] to cover both positive and negative musical emotions. They comprised the first 60 seconds of the seven compositions shown in Table 1. Altogether, these seven pieces were used to induce emotions with different levels of valence and arousal.

TABLE 1
MUSICAL EXCERPTS USED DURING THE STUDY.

| No. | Name | Composer |
|---|---|---|
| 1 | Neynava (Neynava) | Alizadeh |
| 2 | concierto de aranjuez (Adagio) | Rodrigo |
| 3 | Shabe vasl (Rang Shahr Ashoub) | Darvish khan |
| 4 | Eine kleine nacht music(Rondo Allegro) | Mozart |
| 5 | Le Quattro stagioni (La primavera) | Vivaldi |
| 6 | Nocturne Op. 9 (No. 2) | Chopin |
| 7 | Peer Gynt's Suite no.2(Solveigs song) | Grieg |

## 2.3 Experimental protocol

To ensure that all participants were equally familiar with the musical stimuli, two to four days before the experiment the subjects were presented with the stimuli and were instructed to listen to each piece in a calm environment only once. Participants sat in a comfortable chair in dim light. The stimuli were presented via suitable headphones at a comfortable volume which was the same across all subjects. During EEG recordings, the stimuli were played for the participants, and they were instructed to pay attention to the compositions. Right after EEG recordings the musical excerpts were once again played for the participants and while listening to each excerpt, they expressed their continuous emotional appraisals by moving the cursor on a 2-dimensional arousal-valence (A-V) plane. FEELTRACE [37] was used for recording the time-varying participants' ratings of A-V. For further processing, the sampling rate of the recorded A-V data was reduced to 128 Hz. EEG recording of the scalp was carried out using Emotiv EPOC 14-channel EEG wireless recording headset (Emotiv Systems, Inc., San Francisco, CA). This device obtains EEG data with an internal sampling frequency of 2048. The output data was later downsampled to 128 Hz. Electrode placement was according to the 10–20 system (Electro Cap International Inc., Eaton, USA), including positions AF3, F7, F3, FC5, T7, P7, O1, O2, P8, T8, FC6, F4, F8, and AF4 as illustrated in Figure 1.a. During the preparation steps we were cautious about the correct placement of the headset on each subject's head. Common mode reference was used. The portability and easy-to-wear characteristics of the recording device makes it a good candidate for BCI applications [38]. During each recording session participants completed six consecutive trials, with each trial consisting of two 5 s silence intervals, two 5 s white noise intervals and one 60 s stimulus. One trial sequence is depicted in Figure 1.b.

## 2.4 EEG processing

After data acquisition, EEG signals were band-pass filtered with a lower cut off frequency of 2 Hz and upper cut off frequency of 42 Hz to reduce EMG and power line artifacts. Next, independent component analysis (ICA) was performed on the EEG signals. To reject the ICA components a multiple artifact rejection algorithm (MARA) was employed using EEGLAB plug-in MARA [39]. The EEG signals were later transferred to a common average montage. EEG signals were visually inspected to remove the remaining artifacts. EEGLAB toolbox [40] was used for EEG data preprocessing.

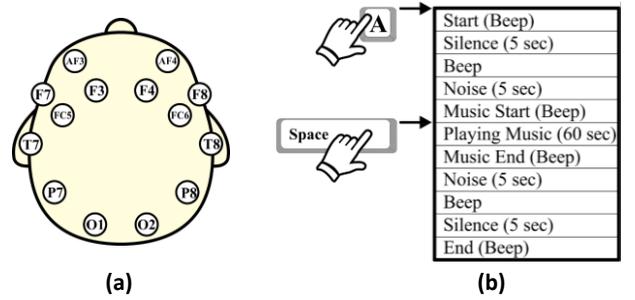

Fig. 1. (a) Channel locations. (b) One trial during the experimental protocol.

Since the emotional appraisal signals, which were considered as behavioral responses, did not contain high-frequency components [41] and also to remove noises corresponding to handshakes during cursor navigations, a moving average window was used to smooth the appraisal data. Figure 2 shows valence and arousal signals corresponding to musical selection (MS) no. 2 (Table 1), evaluated by one of the subjects, along with their smoothed versions.

To estimate emotions from EEG signals, the proposed model was used to find the relation between inputs (time-varying power spectra of EEG signals) and outputs (individual emotion appraisal signals) of the system. To calculate time-varying EEG power spectra Morlet wavelet transform was used. The applied real-valued Morlet wavelet is defined as [42]:

$$\psi(t) = \exp(-\frac{t^2}{2})\cos(5t) \qquad (1)$$

The continuous wavelet transform of a signal $x(t)$, $W_x(t',s)$, will be obtained by convolution of the signal with shifted and scaled versions of the mentioned real Morlet wavelet:

$$W_x(t',s) = \frac{1}{s}\int x(t)\psi(\frac{t-t'}{s}) \qquad (2)$$

where $t'$ and $s$ indicate the shift and scale parameter, respectively. The continuous wavelet transform was computed for all frequency bands. The scale parameter was calculated based on the frequency range of theta (4-8 Hz), alpha (8-13Hz) and beta (13-31Hz) frequency bands to obtain time-varying power spectra of each band. Then the power spectra in every band were averaged over the frequency range in corresponding band to obtain one time-varying signal for every frequency bands. This procedure was repeated for every channel; thus for all channels we produced 42 input candidates (4 frequency bands×14 channels).

## 2.5 Fuzzy Parallel Cascades

One of the commonly used paradigms in nonlinear system identification is the parallel cascade structure [43]. In the current study, we used this structure and combined its concept with a fuzzy inference system to introduce a novel approach, called Fuzzy Parallel Cascades (FPC). Figure 3 shows the structure of the proposed model. Fuzzy logic success is due to its capability in describing system dynamics by simple conditional IF-THEN rules. In most applications, this capability provides a simple solution which requires relatively short computation time. Also, all information and engineering knowledge related to the structure of the system, as well as its optimization, can be employed directly [31, 44]. In this study, we applied a look-up table scheme [45] for rule extraction, which will be described in the next section. Overall, the universal approximation capability [45], interpretability of fuzzy inference systems, the rule extraction ability of look-up table schemes [45] and structure of the parallel cascades paradigm make our proposed method a feasible and transparent identification method for estimating the emotional appraisal response.

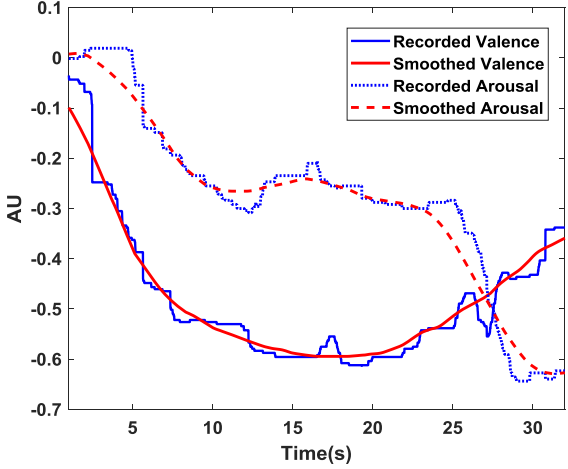

Fig. 2. Time-varying emotion appraisal signals and their corresponding smoothed version for valence and arousal for musical excerpt no. 2 (MS2) by one of the subjects.

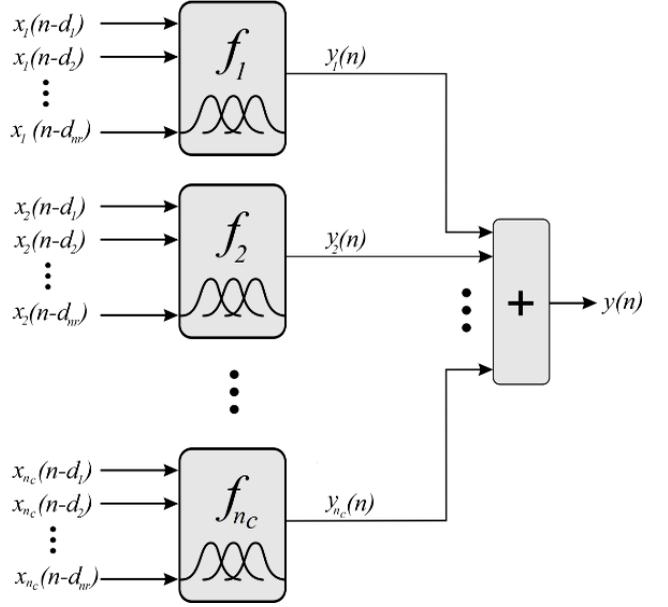

Fig. 3. The proposed model structure. In the *i*-th cascade, xi ,yi are input and outputs, respectively. *d* represents the delay of input, and *fi* is the fuzzy logic-based model developed for the *i*-th cascade.

The proposed model includes $n_c$ cascades with each having a separate set of inputs. If the main input of the system is $x_i(n)$, the inputs of the *i*-th cascade $x_i(n-d_1), x_i(n-d_2), ..., x_i(n-d_{n_r})$ are the delayed versions of $x_i(n)$ where $n_r$ is the number of delays. Each of the previously introduced 32 input candidates can be employed in this step. Since we had 32 input candidates, the maximum number of cascades was limited to 32. Each cascade includes a fuzzy logic-based model represented by $f_i$ which will be described in the following paragraphs. The estimated output of the *i*-th cascade $\hat{y}_i(n)$ is calculated as

$$\hat{y}_i(n) = f_i(x_i(n-d_1), x_i(n-d_2), ..., x_i(n-d_{n_r})) \quad (3)$$

and the residue of the *i*-th cascade is

$$r_i(n) = y_i(n) - \hat{y}_i(n) \quad (4)$$

where $y_i(n)$ is the measured or desired output of the *i*-th cascade. For the first cascade we have the following equation

$$y_1(n) = y(n) \quad (5)$$

where $y(n)$ is the *n*-th sample of preprocessed time-varying appraisal signal and for $i = 2, 3, ..., n_c$ we have

$$y_i(n) = r_{i-1}(n) \quad (6)$$

where $r_{i-1}(n)$ is the residue of the previous cascade. Roughly speaking, this means that by adding a new cascade to the model we intended to estimate the nonlinear dynamics of the system which could not be determined using the already developed cascades.

This procedure is repeated for $n_c$ cascades. Finally, the estimated output of model, $\hat{y}(n)$, is obtained by

$$\hat{y}(n) = \sum \hat{y}_i(n) \quad (7)$$

To avoid overfitting, we imposed a residue criterion ($r_{cr}$) to limit the number of cascades.

$$r_{cr} = \frac{\left\| \left( \sum_{i=1}^{c} \hat{y}_i \right) - y \right\|}{\left\| \left( \sum_{i=1}^{c-1} \hat{y}_i \right) - y \right\|} \quad (8)$$

This criterion represents the error reduction ratio due to adding the *c*-th cascade to the model. The *c*-th cascade is added if $r_{cr}$ for the validation data is less than one, i.e., appending cascades is continued only until the overall system error keeps a decreasing manner. This procedure determines the number of cascades.

Therefore our model consisted of $n_c$ cascades with each having a multi-input/single-output (MISO) fuzzy system with $n_r$ inputs and one output. To simplify, we rename $x_i(n-d_k)$ as $x_{ik}(s)$ where $s = 1, 2, ..., l_s$ with $l_s$ being the length of inputs and output data vectors. For each input vector $\mathbf{x}_{ik}$, $n_{m_x}$ fuzzy sets and for each output vector $\mathbf{y}_i$, $n_{m_y}$ fuzzy sets are defined. Membership functions for the *k*-th delay of the input of the *i*-th cascade, $\mathbf{x}_{ik}$, and output are named $\Gamma_k^j (j = 1, 2, ..., n_{m_x})$ and $\Lambda^l (l = 1, 2, ..., n_{m_y})$, respectively. If we consider three Gaussian membership functions for $\mathbf{x}_{ik}$, the continuous partitioning of it will be as depicted in Figure 4.

By the ordered arrangement of *s*-th samples of output vector $\mathbf{y}_i$ and input vectors $\mathbf{x}_{ik}$, $k = 1, 2, ..., n_r$, an ordered $(n_r + 1)$-tuples, $(\chi_i(s), y_i(s)), s = 1, 2, ..., l_s$, is created where

$$\chi_i(s) = (x_{i1}(s), x_{i2}(s), ..., x_{in_r}(s)) \quad (9)$$

We have

$$\chi_i(s) \in U = \left[ a_{x_1}, b_{x_1} \right] \times ... \times \left[ a_{x_{n_r}}, b_{x_{n_r}} \right] \subset R^{n_r} \quad (10)$$

and
$$y_i(s) \in V = [a_y, b_y] \subset R \quad (11)$$
where $U$ and $V$ are the universal sets of inputs and output, respectively. $\Lambda^l (l=1,2,\ldots,n_{m_y})$ and $\Gamma_k^j (j=1,2,\ldots,n_{m_x})$ fuzzy sets are complete in the corresponding intervals. First, for every $(\chi_i(s), y_i(s)), s=1,2,\ldots,l_s$, membership values of $x_{ik}(s), k=1,2,\ldots,n_r$ in fuzzy sets $\Gamma_k^j (j=1,2,\ldots,n_{m_x})$ and membership values of $y_i(s)$ in fuzzy sets $\Lambda^l (l=1,2,\ldots,n_{m_y})$ are determined, and named $\mu_{\Gamma_k^j}(x_{ik}(s))$ and $\mu_{\Lambda^l}(y_i(s))$, respectively.

Next, for every input, the fuzzy set in which $x_{ik}(s)$ had the largest membership value is determined (and named, e.g. $\Gamma_k^{\tilde{j}}$). A similar procedure is carried out for the output which results in $\Lambda^{\tilde{l}}$. Finally, an IF-THEN fuzzy rule for the $s$-th sample is

$$\text{if } x_{i1}(s) \text{ is } \Gamma_1^{\tilde{j}} \text{ and}\cdots\text{and } x_{in_r}(s) \text{ is } \Gamma_{n_r}^{\tilde{j}} \text{ then } y_i(s) \text{ is } \Lambda^{\tilde{l}} \quad (12)$$

Due to the large number of input-output ($n_r+1$)-tuples, it is possible to have some antithetical rules with similar "*if*" and different "*then*" parts. To solve this problem, a weight is assigned to every produced rule using the following equation:

$$DR = \prod_{k=1}^{n_r} \mu_{\Gamma_k^{\tilde{j}}}(x_{ik}(s)) \mu_{\Lambda^{\tilde{l}}}(y_i(s)) \quad (13)$$

The strongest rule among antithetical rules is chosen, and the remaining rules are removed [45]. By imposing the above fuzzy rules we create $n_c$ fuzzy systems for the realization of the FPC model. Each of these fuzzy systems uses the Mamdani product inference engine. This inference engine is equipped with algebraic product T-norm, max s-norm, and Mamdani product implications. Supposing that the rule extraction process produced M consistent rules, by using the generalized Modus Ponens rule the membership function of the inference output of the $i$-th fuzzy system for $s$-th sample is [45]:

$$\mu_{\hat{\Lambda}}(y_i(s)) = \max_{p=1}^{m}\left(\sup_{\chi_i \in U}\left[\mu_{\hat{\Gamma}}(x_{i1}(s),\ldots,x_{in_r}(s))\right] \times \prod_{k=1}^{n_r} \mu_{\Gamma_k^{\tilde{j}}}(x_{ik}(s))\mu_{\Lambda^{\tilde{l}}}(y_i(s))\right) \quad (14)$$

where $\mu_{\Gamma_k^{\tilde{j}}}(x_{ik}(s))$ and $\mu_{\Lambda^{\tilde{l}}}(y_i(s))$ are Gaussian membership functions

$$\mu_{\Gamma_k^{\tilde{j}}}(x_{ik}(s)) = \exp\left(-\left(\frac{x_{ik}(s) - cx_{ik}^{\tilde{j}}}{\sigma x_{ik}^{\tilde{j}}}\right)^2\right) \quad (15)$$

and

$$\mu_{\Lambda^{\tilde{l}}}(y_i(s)) = \exp\left(-\left(\frac{y_i(s) - cy_i^{\tilde{l}}}{\sigma y_i^{\tilde{l}}}\right)^2\right) \quad (16)$$

$cx_{ik}^{\tilde{j}}$ and $cy_i^{\tilde{l}}$ are centers and $\sigma x_{ik}^{\tilde{j}}$ and $\sigma y_i^{\tilde{l}}$ are standard deviations (SD) of the $k$-th input and output membership function of the $i$-th fuzzy system, respectively. In the last stage by using the center of gravity defuzzification we calculate the numerical output of the $i$-th fuzzy system as follows

$$\hat{y}_i = \frac{\int_V y_i \, \mu_{\hat{\Lambda}}(y_i) dy_i}{\int_V \mu_{\hat{\Lambda}}(y_i) dy_i} \quad (17)$$

Different values for $n_r$ and the number of membership functions are evaluated using repeated experiments to produce the most accurate estimates. The process above resulted in the following values, $n_{m_x} = n_{m_y} = 3$, $n_r = 2$.

During the identification phase, we employed the MI for selection of the appropriate input to each cascade from the pool of inputs. To do so, input candidates were sorted based on their MI with the corresponding output signal. MI shows the amount of information shared between the two variables [46]. For two random variables $x, y$, MI defined as

$$I(X;Y) = \sum_{x,y} p(x,y) \log \frac{p(x,y)}{p(x)p(y)} \quad (18)$$

where $p(x,y)$ is the joint probability function of $x, y$ and $p(x), p(y)$ are marginal probability functions of $x, y$ respectively. In the case of independent $x, y$, MI is zero. Once an input was chosen it was removed from the pool of input candidates and hence not considered in the successive steps. As mentioned earlier, adding cascades was continued until $r_{cr}$ was less than one.

## 3 RESULTS

### 3.1 Behavioral Result

We recorded the valence and arousal values reported by participants on the FEELTRACE screen while listening to 60 s of each musical excerpt (Table 1). A part of the mean of continuous reported valence and arousal of all musical excerpts are illustrated in Figure 6. where the shadowed range indicates standard deviation. The average value of reported valence and arousal for every musical excerpt are also demonstrated in Table 2. According to the average values of reported valence, we categorized MS1, and MS2 as "melancholic", MS3, MS4, and MS5 as "joyful" and MS6 and MS7 as "neutral". Only the valence is considered for music categorization because we have chosen MSs based on the positivity and negativity of emotion which implies valence.

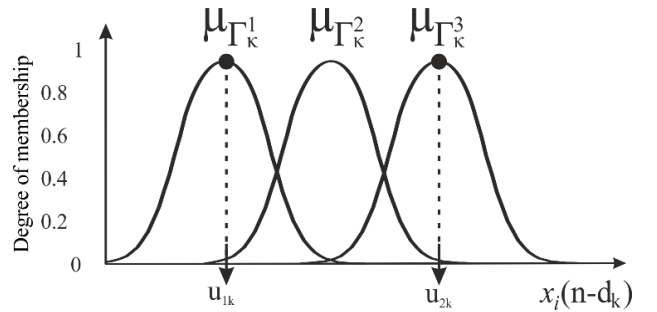

Fig. 4. Three Gaussian membership functions for partitioning the input space of $x_i(n-d_k)$. $i$ and $k$ are the number of cascades and delays, respectively. $\mu_{\Gamma_k^j}$ is the membership function of the input in the j-th fuzzy set $\Gamma_k^j$.

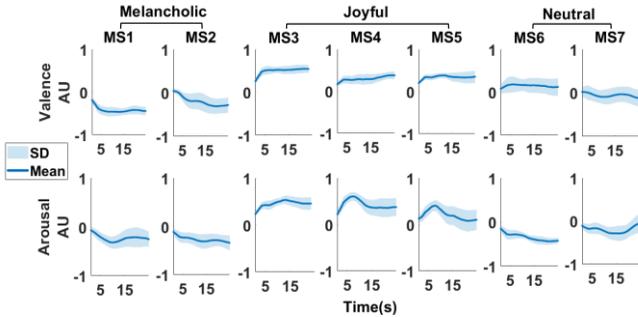

Fig. 5. Mean and standard deviation (SD) signals of valence and arousal reported by participants for each musical excerpt in time. Unit of the y-axis of all plots is AU (arbitrary unit).

To investigate whether emotional ratings among these categories are different, a statistical test on the average ratings between the categories was performed. Since the distribution of data wasn't normal, the Kruskal-Wallis test was applied. The results of the statistical test are demonstrated in Table 3. As can be seen from Table 3 valence and arousal for all categories of melancholic, joyful, and neutral are significantly different (p<0.05).

### 3.2 Mutual Information of Emotion and EEG Features

During the identification process, MI of EEG signals and emotion appraisal was used to select model inputs. In this way, the EEG signals which have higher MI with emotion applied as inputs to estimate emotion. It means that EEG electrodes with higher MI were those who played a more critical role in producing the emotional appraisal signal. Since MI shows the information shared between two signals [46], to find out which EEG electrodes shared more information with emotion signals, mean MI between power spectra of different channels in each frequency band (i.e. theta, alpha, and beta) with A-V signals for different MSs are shown in Figure 6.a and 6.b for valence and arousal respectively. From Figure 6 it seems that MI values of channels located in the frontal region with both valence and arousal are higher than other channels, especially in theta and alpha frequency band. This observation confirmed reports of previous researches in the role of this region in the processing of emotional stimuli [3]. Based on Davidson's frontal brain asymmetry hypothesis, left frontal activation are related to experiencing emotions with positive valence while right frontal activation is associated with negative emotions[47].

TABLE 2
MEAN AND STANDARD DEVIATION (SD) OF VALENCE AND AROUSAL.

| No. | Name | Valence Mean(SD) | Arousal Mean(SD) |
|---|---|---|---|
| 1 | Neynava (Neynava) | -0.43(0.15) | -0.25(0.26) |
| 2 | concierto de aranjuez (Adagio) | -0.21(0.29) | -0.3(0.24) |
| 3 | Shabe vasl (Rang Shahr Ashoub) | 0.51(0.15) | 0.46(0.16) |
| 4 | Eine kleine nacht music(Rondo Allegro) | 0.3(0.18) | 0.43(0.26) |
| 5 | Le Quattro stagioni (La primavera) | 0.34(0.15) | 0.2(0.32) |
| 6 | Nocturne Op. 9 (No. 2) | 0.15(0.19) | -0.35(0.13) |
| 7 | Peer Gynt's Suite no.2(Solveigs song) | -0.061(0.32) | -0.19(0.27) |

TABLE 3
COMPARISON OF VALENCE AND AROUSAL BETWEEN CATEGORIES.

| Category 1 | Category 2 | p-value Valence | p-value Arousal |
|---|---|---|---|
| Melancholic | Joyful | 9.56E-10 | 9.56E-10 |
| | Neutral | 9.56E-10 | 2.05E-07 |
| Joyful | Melancholic | 9.56E-10 | 9.56E-10 |
| | Neutral | 9.56E-10 | 9.56E-10 |
| Neutral | Melancholic | 9.56E-10 | 2.05E-07 |
| | Joyful | 9.56E-10 | 9.56E-10 |

Considering the contribution of the frontal region in emotion processing [3], we are going to investigate whether MI of frontal channels with emotional appraisal signal in melancholic, joyful and neutral MSs are different or not. Moreover, to inquire Davidson's hypothesis, the MI of EEG electrodes located in left and right frontal with emotion is compared in musical excerpts categories (melancholic, joyful and neutral). It is notable that since frontal brain asymmetry hypothesis is stated for valence of emotion, we only performed the analysis on valence signal. In this way as some data did not have a normal distribution, we have used a Friedman followed by multiple comparisons to compare MI of frontal channels with valence signal among melancholic, joyful and neutral MSs in every frequency band. Two factors that affect every MI value are three categories (melancholic, joyful and neutral) and EEG electrodes although our focus is on categories differences. Frontal channels include AF3, F7, F3, FC5, FC6, F4, F8, and AF4 while AF3, F7, F3, FC5 are located in right frontal and the remains are in left frontal region.

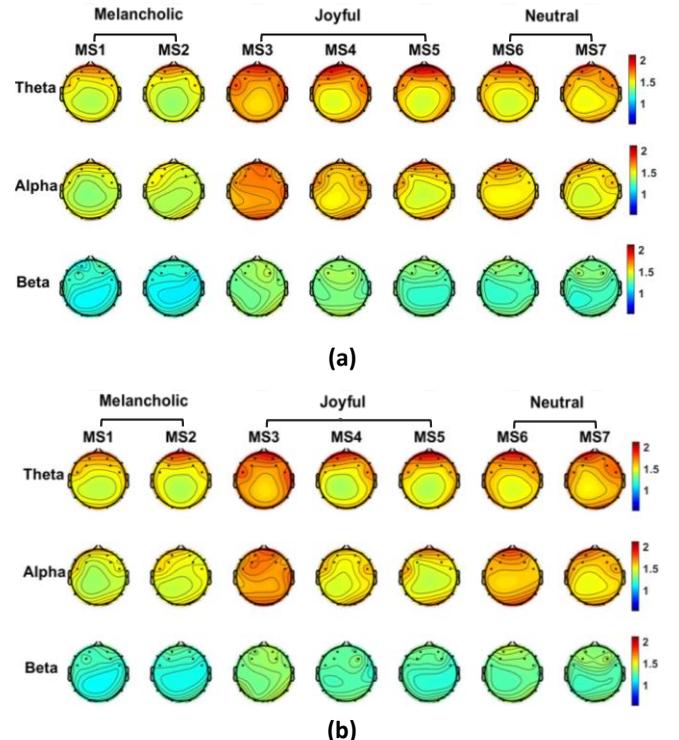

Fig. 6. Mean of mutual information of power spectra of EEG signals with reported valence (a) and arousal (b) for all channels in theta, alpha and beta frequency bands.

Figure 7 displays the mean and standard deviation of MI between right, left, and all frontal EEG and valence of melancholic, joyful and neutral music in theta, alpha and beta bands. The result of the statistical test is also depicted in this figure. The asterisks indicate significant differences (p<0.05 and p<0.001).

As Figure 7 shows in all frequency bands, MI between positive valence (joyful musical excerpts) and EEG of right, left and all frontal channels are significantly higher than MI between negative emotion (melancholic musical excerpts) and EEG. Furthermore, for melancholic and neutral categories, valence of neutral MSs has higher MI with EEG than melancholic MSs while in most of the analysis there aren't any significant differences in MI of studied EEG and valence between joyful and neutral categories. Moreover for every category of MSs, the highest average MI of right, left and all frontal channels with valence are for theta frequency band, and the lowest is for beta frequency band.

### 3.3 Model validation

Performance of the proposed FPC model is assessed by comparing the RMSE of this model with LR, SVR and LSTM-RNN models. LSTM-RNN parameters are selected based on [7] and it has two hidden layers, including an LSTM layer. The learning rate was selected 0.001 with the momentum of 0.9.

In the modeling process of all models, for all musical excerpts and all subjects we divided the corresponding inputs (EEG spectra of different channels) and output (valence or arousal) data vectors into two parts for identification and validation phases. Therefore for every musical excerpt and subject, the model was identified using the data corresponding to the half of each MS and validated by the other part of each MS. The procedure mentioned above was investigated for both valence and arousal dimensions. RMSE was used to evaluate the models' performance. Mean of RMSE of LR, SVR, LSTM-RNN, and FPC, calculated over all subjects for both valence and arousal dimension, are presented in Table 4. It can be seen that FPC obtains the lowest model error for estimation of both valence and arousal in all musical excerpts; however the SVR model leads to highest RMSE in most of the analysis. Moreover in estimation of valence of joyful category and arousal of melancholic categories LR outperformed the LSTM-RNN. The lowest obtained RMSE value is 0.089 which belongs to valence estimation of MS4 by FPC model.

## 4 DISCUSSION

In this study we proposed a fuzzy parallel cascades model to predict the continuous emotional content of musical excerpts with different emotional contents by the time-varying spectral content of EEG. Time-varying spectral of EEG is computed by wavelet transform, and the emotional rating was recorded continuously in two dimensions of valence and arousal. Figure 6 illustrates that the perceived emotion is changing during the time while many previous researchers omit this temporal information by considering the labels for describing emotion [1-3].

The musical excerpts used in the current study to induce emotions are categorized to melancholic, joyful and neutral based on their reported valence. The results of statistical testing reported in Table 3 indicate that valence and also arousal of each category are different which confirms proper categories selection. The FPC model is constructed by parallel cascades which enables the model to improve the output estimation through the cascades. The input of each cascade is electrode in theta, alpha or beta frequency bands and delayed versions of it. The MI of time-varying power spectra of EEG with the model output (valence/arousal) has been used to specify the order of applying inputs to cascades. Inquiring the obtained MI values between EEG of right, left and all frontal channels with valence in three categories of melancholic, joyful and neutral MSs indicates higher MI of EEG with positive valence (joyful MSs) compared to negative valence (melancholic MSs). This observation disaffirms the frontal brain asymmetry hypothesis. Some of the previous studies also reported lack of this asymmetry [21, 48]. Inspecting the difference of obtained MI among studied frequency bands demonstrates that frontal EEG in theta frequency band has more relation with valence than alpha or beta bands. This result is consistent with previous researches which reported the involvement of frontal region in theta frequency band during emotion processing [10].

The performance of the FPC model is compared with LR, SVR, and LSTM-RNN model which were utilized for continuous affect recognition in previous studies [7, 28, 29].

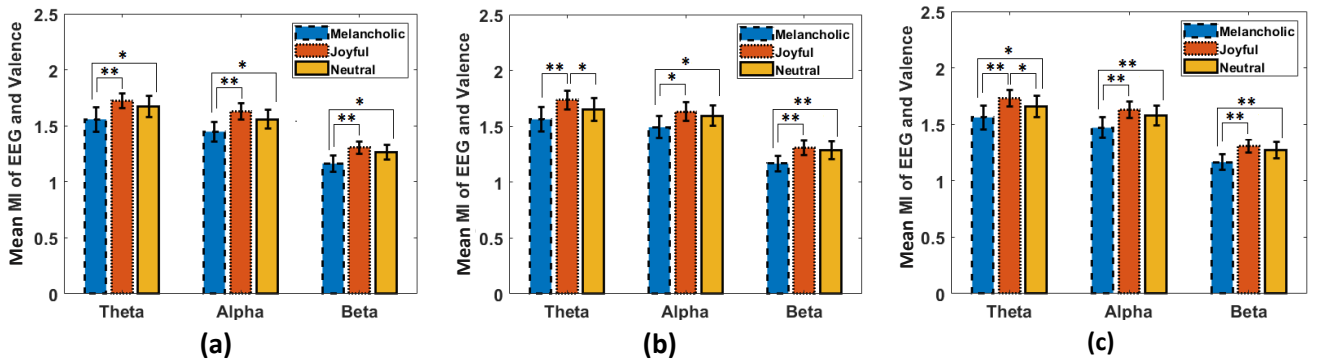

Fig. 7. Mean and standard deviation of mutual information (MI) between EEG of (a) right, (b) left and (c) all frontal channels and valence for melancholic, joyful and neutral music in theta, alpha and beta bands. Significant differences of MI between categories of musical excerpts are shown by asterisks (*: p<0.05 and **: p<0.001).

TABLE 4
RMSE OF MODELS FOR VALENCE AND AROUSAL ESTIMATION.

| Music | V/A | LR Mean(SD) | SVR Mean(SD) | LSTM-RNN Mean(SD) | FPC Mean(SD) |
|---|---|---|---|---|---|
| MS1 | Valence | 0.181(0.13) | 0.191(0.13) | 0.16(0.11) | 0.103(0.088) |
| | Arousal | 0.219(0.13) | 0.234(0.15) | 0.22(0.16) | 0.11(0.068) |
| MS2 | Valence | 0.249(0.14) | 0.269(0.16) | 0.234(0.15) | 0.166(0.12) |
| | Arousal | 0.201(0.13) | 0.209(0.14) | 0.206(0.14) | 0.108(0.059) |
| MS3 | Valence | 0.139(0.1) | 0.154(0.1) | 0.14(0.11) | 0.106(0.044) |
| | Arousal | 0.17(0.09) | 0.19(0.098) | 0.171(0.11) | 0.109(0.053) |
| MS4 | Valence | 0.124(0.06) | 0.129(0.07) | 0.133(0.075) | 0.0895(0.03) |
| | Arousal | 0.215(0.23) | 0.228(0.23) | 0.217(0.23) | 0.142(0.099) |
| MS5 | Valence | 0.161(0.15) | 0.172(0.16) | 0.163(0.14) | 0.114(0.088) |
| | Arousal | 0.236(0.17) | 0.234(0.17) | 0.254(0.21) | 0.171(0.12) |
| MS6 | Valence | 0.223(0.14) | 0.257(0.13) | 0.177(0.15) | 0.155(0.14) |
| | Arousal | 0.195(0.08) | 0.213(0.12) | 0.183(0.081) | 0.114(0.065) |
| MS7 | Valence | 0.171(0.12) | 0.182(0.12) | 0.173(0.13) | 0.097(0.059) |
| | Arousal | 0.254(0.2) | 0.269(0.23) | 0.255(0.27) | 0.129(0.088) |

The lower RMSE of the FPC model demonstrates superiority of this model in emotion prediction in comparison to other studied models. This higher performance of FPC model can be related to two main characteristics of it. Firstly the parallel cascade structure which is appropriate for identification of systems with high-order nonlinearities [43] such as EEG. Secondly the applied fuzzy-based system which enables us to describe system dynamics by simple rules and apply expert knowledge to optimize the performance of the system [31, 44].

It is notable that our analysis is based on continuous appraisal and the identification and validation of all applied models are by two different parts of the data of an individual during listening to one musical excerpt. Therefore if emotional content of these two parts of MS were different, the estimation error will be increased.

There are some points in the current study that can be considered in future researches.

Evaluation of continuous affect recognition is an open research problem [49]. The most commonly used measures for evaluation of continuous emotion recognition are RMSE and the Pearson correlation coefficient[49]. However by considering the subjective nature of emotional responses, applying common error based criteria for evaluating the estimated time-varying emotional signal is not plausible. Therefore a new criterion can be developed based on this assumption that if the estimated emotion signal was close to measured emotion, the estimator has acceptable performance and there is no need for the estimated and measured emotion signals to be precisely the same. This definition can be implemented by simple fuzzy rules to have a fuzzy measure for evaluating continuous emotion recognition in future studies.

Moreover, we have identified and validated the models based on different parts of the data of the same participant. This method of model developing may not satisfied generalization thoroughly, as it is one of the important characteristics required for the models to be applicable in a real-life scenario. Therefore one of the other points that can be studied for further analysis is that for the sake of both generality and considering time variable emotional appraisal of music, the model can be identified by one individual and test with others in future works. In this situation, it should be considered that the emotional response of different participants can be affected by their personal parameters such as mood, fatigue, etc [7].

## 5 CONCLUSION

Using proposed fuzzy parallel cascades approach, continuous emotional appraisal in two valence and arousal dimensions was estimated using time-varying EEG power in three frequency bands, namely theta, alpha, and beta. Comparing the RMSE of proposed model with LR, SVR and LSTM-RNN models reveals supremacy of FPC model. The higher MI of time-varying power spectra of frontal EEG in theta band with the valence confirmed the significant role of frontal channels in theta frequency band in emotion recognition reported in former researches. The analysis of MI of right and left frontal EEG electrodes with valence of melancholic and joyful musical excerpts did not indicate laterality in frontal brain in emotion processing. As the proposed model is subjective to generalize it and develop a practical tool, the model would be established on a large database, which perhaps includes subjects with different musical preferences and expertise.

### FUNDING SOURCES

This research did not receive any specific grant from funding agencies in the public, commercial, or not-for-profit sectors.

### REFERENCES

[1] P. C. Petrantonakis and L. J. Hadjileontiadis, "Emotion recognition from EEG using higher order crossings," *Information Technology in Biomedicine, IEEE Transactions on*, vol. 14, no. 2, pp. 186-197, 2010.

[2] A. M. Bhatti, M. Majid, S. M. Anwar, and B. Khan, "Human emotion recognition and analysis in response to audio music using brain signals," *Computers in Human Behavior*, vol. 65, pp. 267-275, 2016.

[3] Y.-P. Lin *et al.*, "EEG-based emotion recognition in music listening," *Biomedical Engineering, IEEE Transactions on*, vol. 57, no. 7, pp. 1798-1806, 2010.

[4] G. Balasubramanian, A. Kanagasabai, J. Mohan, and N. G. Seshadri, "Music induced emotion using wavelet packet decomposition—An EEG study," *Biomedical Signal Processing and Control*, vol. 42, pp. 115-128, 2018.

[5] Y. Ding, X. Hu, Z. Xia, Y.-J. Liu, and D. Zhang, "Inter-brain EEG Feature Extraction and Analysis for Continuous Implicit Emotion Tagging during Video Watching," *IEEE Transactions on Affective Computing*, 2018.

[6] F. Noroozi, M. Marjanovic, A. Njegus, S. Escalera, and G.

Anbarjafari, "Audio-visual emotion recognition in video clips," *IEEE Transactions on Affective Computing,* vol. 10, no. 1, pp. 60-75, 2017.

[7] M. Soleymani, S. Asghari-Esfeden, Y. Fu, and M. Pantic, "Analysis of EEG signals and facial expressions for continuous emotion detection," *IEEE Transactions on Affective Computing,* vol. 7, no. 1, pp. 17-28, 2016.

[8] S. Koelsch, *Brain and music*. John Wiley & Sons, 2012.

[9] T. Eerola and J. K. Vuoskoski, "A comparison of the discrete and dimensional models of emotion in music.," *Psychology of Music.,* vol. 39, no. 1, pp. 18-49, 2011.

[10] D. Sammler, M. Grigutsch, T. Fritz, and S. Koelsch, "Music and emotion: electrophysiological correlates of the processing of pleasant and unpleasant music," *Psychophysiology,* vol. 44, no. 2, pp. 293-304, 2007.

[11] L.-O. Lundqvist, F. Carlsson, P. Hilmersson, and P. Juslin, "Emotional responses to music: experience, expression, and physiology," *Psychology of music,* 2008.

[12] A. J. Blood and R. J. Zatorre, "Intensely pleasurable responses to music correlate with activity in brain regions implicated in reward and emotion," *Proceedings of the National Academy of Sciences,* vol. 98, no. 20, pp. 11818-11823, 2001.

[13] K. Mueller *et al.*, "Investigating the dynamics of the brain response to music: A central role of the ventral striatum/nucleus accumbens," *NeuroImage,* vol. 116, pp. 68-79, 2015.

[14] S. Moghimi, A. Kushki, S. Power, A. M. Guerguerian, and T. Chau, "Automatic detection of a prefrontal cortical response to emotionally rated music using multi-channel near-infrared spectroscopy," *Journal of neural engineering,* vol. 9, no. 2, p. 026022, 2012.

[15] S. M. Alarcao and M. J. Fonseca, "Emotions recognition using EEG signals: A survey," *IEEE Transactions on Affective Computing,* 2017.

[16] M. Balconi, E. Grippa, and M. E. Vanutelli, "What hemodynamic (fNIRS), electrophysiological (EEG) and autonomic integrated measures can tell us about emotional processing," *Brain and cognition,* vol. 95, pp. 67-76, 2015.

[17] W. Zheng, "Multichannel EEG-Based Emotion Recognition via Group Sparse Canonical Correlation Analysis," *IEEE Transactions on Cognitive and Developmental Systems,* 2016.

[18] P. Ozel, A. Akan, and B. Yilmaz, "Synchrosqueezing transform based feature extraction from EEG signals for emotional state prediction," *Biomedical Signal Processing and Control,* vol. 52, pp. 152-161, 2019.

[19] F. Hasanzadeh and S. Moghimi, "Emotion estimation during listening to music by EEG signal and applying NARX model and genetic algorithm," presented at the National Conference of Technology, Energy & Data on Electrical & Computer Engineering, 2015.

[20] S. Koelstra and I. Patras, "Fusion of facial expressions and EEG for implicit affective tagging," *Image and Vision Computing,* vol. 31, no. 2, pp. 164-174, 2013.

[21] F. Hasanzadeh, H. Shahabi, S. Moghimi, and A. Moghimi, "EEG investigation of the effective brain networks for recognizing musical emotions," *Signal and Data Processing,* Research vol. 12, no. 2, pp. 41-54, 2015.

[22] H. Shahabi and S. Moghimi, "Toward automatic detection of brain responses to emotional music through analysis of EEG effective connectivity," *Computers in Human Behavior,* vol. 58, pp. 231-239, 2016.

[23] P. Li *et al.*, "EEG based emotion recognition by combining functional connectivity network and local activations," *IEEE Transactions on Biomedical Engineering,* 2019.

[24] T. Song, W. Zheng, P. Song, and Z. Cui, "EEG emotion recognition using dynamical graph convolutional neural networks," *IEEE Transactions on Affective Computing,* 2018.

[25] J. A. Russell, "A circumplex model of affect.," *Journal of personality and social psychology,* vol. 39, no. 6, pp. 1161-1178, 1980.

[26] R. E. Thayer, "The biophysiology of mood and arousal," ed: New York: Oxford University Press, 1989.

[27] W. Liu, L. Zhang, D. Tao, and J. Cheng, "Reinforcement online learning for emotion prediction by using physiological signals," *Pattern Recognition Letters,* vol. 107, pp. 123-130, 2018.

[28] M. A. Nicolaou, H. Gunes, and M. Pantic, "Continuous prediction of spontaneous affect from multiple cues and modalities in valence-arousal space," *IEEE Transactions on Affective Computing,* vol. 2, no. 2, pp. 92-105, 2011.

[29] Q. Mao, Q. Zhu, Q. Rao, H. Jia, and S. Luo, "Learning Hierarchical Emotion Context for Continuous Dimensional Emotion Recognition From Video Sequences," *IEEE Access,* vol. 7, pp. 62894-62903, 2019.

[30] V. Z. Marmarelis, *Nonlinear dynamic modeling of physiological systems*. John Wiley & Sons, 2004.

[31] M. Annabestani and N. Naghavi, "Nonlinear identification of IPMC actuators based on ANFIS–NARX paradigm," *Sensors and Actuators A: Physical,* vol. 209, pp. 140–148, 2014.

[32] M. Annabestani and M. Saadatmand-Tarzjan, "A New Threshold Selection Method Based on Fuzzy Expert Systems for Separating Text from the Background of Document Images," *Iranian Journal of Science and Technology, Transactions of Electrical Engineering,* journal article vol. 43, no. 1, pp. 219-231, July 01 2019.

[33] F. Hasanzadeh and F. Faradji, "An ICA Algorithm Based on a Fuzzy Non-Gaussianity Measure," presented at the 1st Conference on New Research Achievements in Electrical and Computer Engineering, 2016.

[34] M. Annabestani, A. Rowhanimanesh, A. Mizani, and A. Rezaei, "Fuzzy descriptive evaluation system: real, complete and fair evaluation of students," *Soft Computing,* journal article May 20 2019.

[35] M. Annabestani, N. Naghavi, and M. Maymandi-Nejad, "From modeling to implementation of a method for restraining back relaxation in ionic polymer–metal composite soft actuators," *Journal of intelligent material systems and structures,* vol. 29, no. 15, pp. 3124-3135, 2018.

[36] I. Peretz, L. Gagnon, and B. Bouchard, "Music and emotion: perceptual determinants, immediacy, and

isolation after brain damage," *Cognition,* vol. 68, no. 2, pp. 111-141, 1998.

[37] R. Cowie, E. Douglas-Cowie, S. Savvidou*, E. McMahon, M. Sawey, and M. Schröder, "'FEELTRACE': An instrument for recording perceived emotion in real time," in *ISCA tutorial and research workshop (ITRW) on speech and emotion*, 2000.

[38] S. K. Hadjidimitriou and L. J. Hadjileontiadis, "Toward an EEG-based recognition of music liking using time-frequency analysis," *Biomedical Engineering, IEEE Transactions on,* vol. 59, no. 12, pp. 3498-3510, 2012.

[39] I. Winkler, S. Haufe, and M. Tangermann, "Automatic classification of artifactual ICA-components for artifact removal in EEG signals," *Behavioral and Brain Functions,* vol. 7, no. 1, p. 30, 2011.

[40] A. Delorme and S. Makeig, "EEGLAB: an open source toolbox for analysis of single-trial EEG dynamics including independent component analysis," *Journal of neuroscience methods,* vol. 134, no. 1, pp. 9-21, 2004.

[41] F. Upham, "Quantifying the temporal dynamics of music listening: a critical investigation of analysis techniques for collections of continuous responses to music.," McGill University, 2011.

[42] S. Mallat, *A wavelet tour of signal processing*. Elsevier, 1999.

[43] M. J. Korenberg, "Parallel cascade identification and kernel estimation for nonlinear systems," *Annals of biomedical engineering,* vol. 19, no. 4, pp. 429-455, 1991.

[44] M. Annabestani and N. Naghavi, "Nonuniform deformation and curvature identification of ionic polymer metal composite actuators," *Journal of Intelligent Material Systems and Structures,* pp. 1–17, 2014.

[45] L.-X. Wang, *A course in fuzzy systems*. Prentice-Hall press, USA, 1999.

[46] J. Seok and Y. S. Kang, "Mutual information between discrete variables with many categories using recursive adaptive partitioning," *Scientific reports,* vol. 5, p. 10981, 2015.

[47] R. J. Davidson, "Cerebral asymmetry and emotion: Conceptual and methodological conundrums," *Cognition & Emotion,* vol. 7, no. 1, pp. 115-138, 1993.

[48] D. Hagemann, "Individual differences in anterior EEG asymmetry: methodological problems and solutions," *Biological psychology,* vol. 67, no. 1-2, pp. 157-182, 2004.

[49] H. Gunes and B. Schuller, "Categorical and dimensional affect analysis in continuous input: Current trends and future directions," *Image and Vision Computing,* vol. 31, no. 2, pp. 120-136, 2013.